\documentclass[letterpaper, twocolumn,superscriptaddress, showpacs,preprintnumbers,amsmath,amssymb] {revtex4}
\usepackage{graphicx}
\usepackage{dcolumn}
\usepackage{bm}
\usepackage{hyperref}
\hypersetup{
    colorlinks,
    citecolor=red,
    filecolor=blue,
    linkcolor=blue,
    urlcolor=blue
}

\begin{document}

\title{Superdiffusive nonequilibrium motion of an impurity in a Fermi sea}

\author{Hyungwon Kim}
\affiliation{Physics Department, Princeton University, Princeton, NJ 08544, USA}

\author{David A. Huse}
\affiliation{Physics Department, Princeton University, Princeton, NJ 08544, USA}

\begin{abstract}
We treat the nonequilibrium motion of a single impurity atom in a low-temperature single-species Fermi sea,
interacting via a contact interaction.
In the nonequilibrium regime, the impurity does a superdiffusive geometric random walk where the typical distance
traveled grows with time as $\sim t^{d/(d+1)}$ for the $d$-dimensional system with $d\geq 2$.  For nonzero temperature $T$, this crosses
over to diffusive motion at long times with diffusivity $D\sim T^{-(d-1)/2}$.
These results apply also to a nonzero concentration of impurity atoms as long as they remain dilute and nondegenerate.
\end{abstract}


\maketitle
In condensed matter physics, the behavior of a single impurity immersed in a sea of majority particles has been one of the simplest many-body problems
and attracted much attention (see, e.g. \cite{mitra,devreese}).
Especially, transport of an impurity in Bosonic \cite{abe} and Fermionic \cite{clark,schappert} quantum liquids has been studied since a few decades ago.
Recent realizations of ultra-cold polarized two-component atomic Fermi gases \cite{partridge1,zwierlein,shin1,partridge2,shin2,shin3,nascimbene},
with their remarkable controllability over parameters such as interaction strength and individual populations,
have made it possible to directly access this type of quantum many-body system.
Meanwhile, there have been a number of theoretical works on the single impurity problem in
an ultracold Fermi gas \cite{chevy,lobo,combescot,prokofev,mora},
investigating the equilibrium and near-equilibrium properties.
Quite naturally, transport phenomena
have also been studied \cite{bruun}.
More recently, transport experiments of an ultracold mass-balanced polarized Fermi gas \cite{sommer2}
and an ultracold mass-imbalanced mixture with unequal populations \cite{trenkwalder}
opened new opportunities to directly investigate nonequilibrium and dynamic properties of population-imbalanced Fermi systems.

In this paper, we discuss nonequilibrium transport in the high polarization and low temperature regime,
without restricting ourselves within small deviations from equilibrium.
As a limiting case of high polarization and low temperature,
we first consider a single minority atom moving in a zero-temperature Fermi sea of majority atoms.
We find that the impurity
does a superdiffusive geometric random walk, where the time between collisions grows
in proportion to time, and the impurity loses a fraction of order one of its energy in each collision.
In $d$ dimensions ($d\geq 2$) the typical distance traveled grows as $\sim t^{d/(d+1)}$.

As is conventional, we call the majority species ``up" $\uparrow$ and the minority (impurity) species ``down" $\downarrow$.
Note that in the regimes we study in this paper, the statistics of the minority atoms do not enter, so they
may equally well be bosons or fermions.
Assume the minority atom is initially near the origin in real space, with some probability distribution
of its momentum ${\bf Q_{\downarrow}}$ with
$Q_{\downarrow}\ll k_{F\uparrow}$, where $k_{F\uparrow}$ is the Fermi momentum of the majority
Fermi sea.  The minority atom is ``dressed''
either as a polaron or as a molecule, with effective mass $m^*$ and thus
energy $E_{\downarrow}\approx\frac{\hbar^2 Q_{\downarrow}^2}{2m^*}$; we choose the
rest energy of the dressed minority atom as its zero of energy.
We assume that $E_{\downarrow}$ is low enough so that no internal excitations of the
polaron or molecule are possible.
Note that Fermi polarons and bosonic molecules in this high-polarization limit
were experimentally studied in Ref. \cite{schirotzek}.

We will treat the majority atoms as noninteracting, although the results we obtain appear
to remain qualitatively correct even if the majority atoms do weakly interact with one another,
as long as the majority atoms remain a Fermi liquid.  In the latter case, the minority atom produces
quasiparticles and quasiholes when it scatters from the majority Fermi liquid.
We will first treat the case of a three-dimensional Fermi gas, returning to general $d$ later.

This dressed minority atom can scatter from a majority
atom with momentum $q_{\uparrow}$ slightly below $k_{F\uparrow}$ and thus emit a majority particle-hole pair,
which lowers the energy of the minority atom.  Here we are considering such scattering events that occur
``on-shell''; the virtual particle-hole pairs that dress the polaron or molecule are assumed to have already
been included and have renormalized the rest energy and the effective mass of the dressed impurity atom.
After the scattering event, the
emitted majority particle and hole move away ballistically at the majority Fermi velocity.
In such a scattering event the dressed minority atom typically loses
a fraction of order one of its energy.  The rate of scattering
is proportional to $E^2_{\downarrow}$ for $d=3$, as we derive below.
As a result the typical energy evolves with time as
\begin{equation}\label{rate}
\frac{1}{E_{\downarrow}}\frac{dE_{\downarrow}}{dt}\sim -E_{\downarrow}^2~,
\end{equation}
which results in $E_{\downarrow}(t)\sim t^{-1/2}$ and typical speed
$v \sim Q_{\downarrow}\sim t^{-1/4}$.  Thus the minority atom does a
superdiffusive geometric random walk, where the typical time $\tau$ between scattering events
grows as $\tau\sim t$ and the mean free path $\ell$
grows as $\ell\sim v\tau\sim t^{3/4}$.
The last step of this walk typically gives the dominant contribution to the total distance traveled, so
the latter also grows as $\sim t^{3/4}$.
The number of majority particles and holes produced and the number
of steps in this unusual random walk grow only as $\sim \log{t}$.

In order to understand this behavior more quantitatively, we consider the time-dependence of the
momentum distribution
$f({\bf Q_\downarrow}, t)$ of the dressed minority atom.
$f({\bf Q_\downarrow},t)$ has a dimension of (length)$^d$ and is normalized to unity when integrated over wavevectors ${\bf Q_{\downarrow}}$.
We essentially treat the minority atom classically.  Although its initial state may be a quantum-coherent wavepacket and thus of low entropy, as it produces particle-hole
pairs it becomes more and more highly entangled with these excitations
that it has produced in the majority Fermi sea.
As a result, at long times its reduced density matrix is a mixed state, any initial coherence is transferred to the Fermi sea,
and the volume occupied by the minority atom in its
position-momentum configuration space grows as $\sim(\ell Q_{\downarrow})^3\sim t^{3/2}$, so its entropy grows as $\sim\frac{3}{2}k_B\log{t}$.

A single minority atom in an infinite system does not disturb the
distribution of the majority atoms, which are thus assumed to remain in a
Fermi-Dirac distribution at all times.  We now consider both the case of a zero-temperature majority Fermi sea,
as well as low nonzero temperature $T\ll T_{F\uparrow}$ where the superdiffusive behavior discussed above
crosses over to diffusion at long time as the minority atom thermally equilibrates with the Fermi sea.
The scattering process is a
majority atom with momentum ${\bf q_{\uparrow}}$ scattering from
the dressed minority atom with momentum ${\bf Q_{\downarrow}}$ in to solid angle $d\Omega$ in the rest frame of the Fermi sea,
resulting in momenta ${\bf k_{\uparrow}}$ and ${\bf Q'_{\downarrow}}$ respectively, and the reverse of this process.
The resulting time-dependence of the minority atom's momentum distribution is
\begin{eqnarray}\label{Boltzmann_Eq}
\frac{df({\bf Q_\downarrow}, t)}{dt} &=& \int \frac{d^3 {\bf q_{\uparrow}}}{(2\pi)^3} v_r \frac{d\sigma}{d\Omega}d\Omega
[n(\epsilon_{k_{\uparrow}})f({\bf Q'_{\downarrow}},t)(1 - n(\epsilon_{q_{\uparrow}}))\nonumber\\
 &&\quad- n(\epsilon_{q_{\uparrow}})f({\bf Q_{\downarrow}},t)(1 - n(\epsilon_{k_{\uparrow}}))] ~,
\end{eqnarray}
where $v_r$ is the speed of the relative motion and $n(\epsilon_k)$ is the Fermi-Dirac distribution with $\epsilon_k=\frac{\hbar^2 k^2}{2m}$ the energy of a majority atom.

In the low energy $Q_\downarrow \ll k_{F\uparrow}$ and low temperature $k_BT \ll \epsilon_{F\uparrow}$ limits, the scattering is all at
relative momenta near $k_{F\uparrow}$ with small momentum transfer
$|{\bf Q'_{\downarrow}-Q_{\downarrow}}|\ll k_{F\uparrow}$. In these limits, $v_r\approx\frac{\hbar k_{F\uparrow}}{m}$ and $\frac{d\sigma}{d\Omega}$ are essentially constant
and may be taken outside of the integral.  The differential cross-section is dependent on the internal structure of the polaron or molecule, for which there
are only approximate results, so we will just take it as an input parameter to our results.
Near unitarity $k_{F\uparrow}|a| > 1$ and $\frac{d\sigma}{d\Omega}\sim k^{-2}_{F\uparrow}$, while away from unitarity $k_{F\uparrow}|a| < 1$ and $\frac{d\sigma}{d\Omega}\sim a^2$, where
$a$ is the bare $s$-wave scattering length.

\begin{figure}
\includegraphics[width=3.00in]{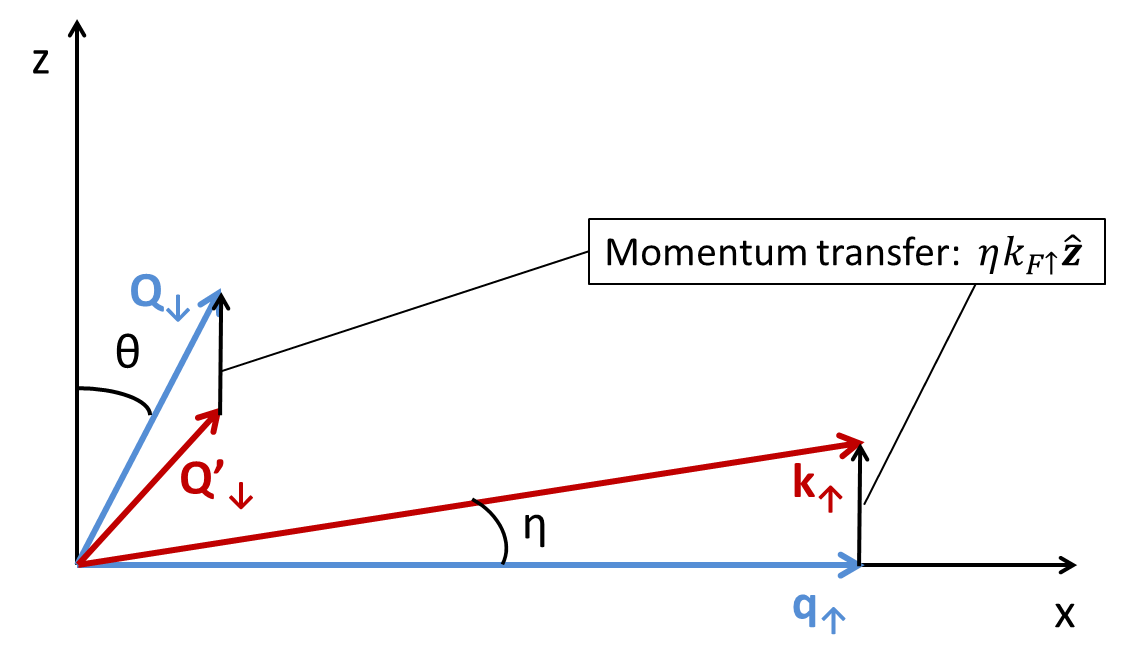}
\centering
\caption{(Color online) Momentum space configuration of the four momenta described in the text. The lengths of vectors are not drawn to scale.}
\label{scheme}
\end{figure}

To scale the dynamics, we assume that distribution of ${\bf Q_{\downarrow}}$ has a characteristic momentum $Q$.  Near equilibrium
$\frac{\hbar^2Q^2}{2m^*}\sim k_BT$, while for a ``hot'', nonequilibrium state $\frac{\hbar^2Q^2}{2m^*}\gg k_BT$.  In either case, the energy
transfer in the scattering is of order $\frac{\hbar^2Q^2}{2m^*}\ll \epsilon_{F\uparrow}$.  Let's put the $x$ axis along ${\bf q_{\uparrow}}$, so
${\bf q_{\uparrow}}={\bf\hat x}(k_{F\uparrow}+q)$.  The momentum transfer is small, so the angle $\eta$ between ${\bf q_{\uparrow}}$ and ${\bf k_{\uparrow}}$
is small $\eta\sim\frac{Q}{k_{F\uparrow}}\ll 1$; we define $\eta$ to be nonnegative.  Put the $z$ axis so ${\bf k_{\uparrow}}=({\bf \hat x\cos{\eta}+\hat z\eta})(k_{F\uparrow}+k)$.
Due to $k_BT\ll\epsilon_{F\uparrow}$ and the small energy transfer, $|q| \sim |k| \sim \frac{mQ^2}{m^*k_{F\uparrow}}$.  Using these axes to set up
spherical polar angles, ${\bf Q_{\downarrow}}$ points along angles $\theta$, $\phi$, where $\theta$ is the angle from the $z$ axis.
With these coordinates, to the precision needed in this
low energy limit, the momentum transfer is $({\bf Q_{\downarrow}}-{\bf Q'_{\downarrow})=\hat z}\eta k_{F\uparrow}$.
Fig.\ref{scheme} schematically shows the configuration of the four momenta and the axes.
We scale the (small) momentum transfer, defining $\gamma$
so $\gamma Q=\eta k_{F\uparrow}$.  We also set the zero of energy for the majority atoms at their Fermi energy, and scale the (small) distances $q$ and $k$ from the
Fermi surface, defining $p$, so that $\epsilon_{q_{\uparrow}}=\frac{\hbar^2k_{F\uparrow}q}{m}=\frac{\hbar^2Q^2}{m^*}p$.  With all these definitions and
scalings, we obtain
\begin{eqnarray}\label{Scaled Boltzmann_Eq}
&&\frac{df({\bf Q_\downarrow}, t)}{dt} \approx \frac{\hbar Q^4}{(2\pi)^2m^*}\frac{d\sigma}{d\Omega} \int d\phi d(\cos\theta) dp \gamma d\gamma \times\nonumber\\
&&[n(\epsilon_{k_{\uparrow}})f({\bf Q'_{\downarrow}},t)(1 - n(\epsilon_{q_{\uparrow}}))
- n(\epsilon_{q_{\uparrow}})f({\bf Q_{\downarrow}},t)(1 - n(\epsilon_{k_{\uparrow}}))] \nonumber\\
\end{eqnarray}
with $\epsilon_{k_{\uparrow}}=\epsilon_{q_{\uparrow}}+\frac{\hbar^2Q^2}{2m^*}(2\gamma\frac{Q_{\downarrow}}{Q}\cos{\theta}-\gamma^2)$.
Now the integration measures are dimensionless,
so the scaling of the scattering rate $\sim Q^4$ is explicit.  Note that the
result does not depend directly on the majority density, although near unitarity there is a dependence via the differential
cross-section.

If the distribution is isotropic in momentum space, $f(Q_{\downarrow},t)$, then it remains isotropic.  If it starts anisotropic, the
dynamics brings it asymptotically to isotropy at long times.  Assuming isotropy, the integral over $\phi$ can be done, and the integrals over
$\theta$ and $\gamma$ can be exchanged for an integral over $Q'_{\downarrow}$ by multiplying
$\int dQ_\downarrow' \delta(Q'_\downarrow - \sqrt{Q_\downarrow^2 + \gamma^2 Q^2 - 2Q_\downarrow Q \gamma \cos\theta})$, giving
\begin{eqnarray}\label{Isotropic Boltzmann_Eq}
&&\frac{df(Q_\downarrow, t)}{dt} \approx \frac{\hbar Q^2}{\pi m^*}\frac{d\sigma}{d\Omega}
\int dQ'_{\downarrow} dp \frac{Q'_{\downarrow}}{Q_{\downarrow}} {\rm min}\{Q_{\downarrow} , Q'_{\downarrow}\} \times\nonumber\\
&&[n(\epsilon_{k_{\uparrow}})f(Q'_{\downarrow},t)(1 - n(\epsilon_{q_{\uparrow}}))
- n(\epsilon_{q_{\uparrow}})f(Q_{\downarrow},t)(1 - n(\epsilon_{k_{\uparrow}}))]. \nonumber\\
\end{eqnarray}
At zero temperature, the integral over $p$ is readily done, and for the loss term also the integral over $Q'_{\downarrow}$, giving
\begin{eqnarray}\label{Isotropic Boltzmann_Eq}
&&\frac{df(Q_\downarrow, t)}{dt} \approx \frac{\hbar}{2\pi m^*}\frac{d\sigma}{d\Omega} \times\nonumber\\
&&\left[(\int_{Q_{\downarrow}}^{\infty} dQ'_{\downarrow} Q'_{\downarrow}(Q^{'2}_{\downarrow}-Q_{\downarrow}^2)f(Q'_{\downarrow},t))
- \frac{2}{15}Q^4_{\downarrow}f(Q_{\downarrow},t)\right].  \nonumber\\
\end{eqnarray}
Thus we see that at zero temperature, the total rate of scattering of a minority atom with momentum $Q_{\downarrow}$ by
producing a majority particle-hole pair and going to any lower energy is $\frac{\hbar}{15\pi m^*}\frac{d\sigma}{d\Omega}Q^4_{\downarrow}$.

This zero-temperature dynamics has a superdiffusive long-time scaling form in terms of a scaled momentum
\begin{equation}
s=(At)^{1/4}Q_{\downarrow}
\end{equation}
and a scaling function $g(s)$, with
\begin{equation}
f(Q_{\downarrow},t)=(At)^{3/4}g(s) ~.
\end{equation}
With $A=\frac{\hbar}{2\pi m^*}\frac{d\sigma}{d\Omega}$, the scaling function satisfies the integro-differential equation:
\begin{eqnarray}
&&\frac{3}{4}g(s) + \frac{s}{4}\frac{dg(s)}{ds}\nonumber\\
&=& - \frac{2}{15}s^{4}g(s) + \int^{\infty}_{s}ds's'(s'^2-s^2)g(s')~.
\label{masater_eq}
\end{eqnarray}
The resulting $g(s)$ appears to be smooth and of order one at small $s$.  Its large-$s$ asymptotics are:
\begin{equation}
\frac{s}{4}\frac{dg(s)}{ds} \approx -\frac{2}{15}s^{4}g(s) \Rightarrow g(s) \sim \exp\left[-\frac{2}{15}s^{4}\right] ~.
\end{equation}

A small ``cloud'' of minority atoms in a zero temperature Fermi sea will also spread in this same
superdiffusive fashion if it is effectively nondegenerate and noninteracting.  In order for the minority atoms
to be effectively noninteracting, the majority particles and holes
produced must leave the cloud without additional scattering, so the majority mean free path within the cloud must
be larger than the cloud.  If the minority cloud is initially too dense and/or too ``hot'' for this to be true, it will
rapidly expand and cool and cross over to this superdiffusive regime of behavior at later times.

If the majority Fermi sea is not at zero temperature, the superdiffusive behavior will continue until the minority
atom cools down to near equilibrium, where $\frac{\hbar^2Q^2}{m^*}\sim k_BT$.  After that it will move diffusively, with
typical speed $v\sim\sqrt{\frac{k_BT}{m^*}}$ and scattering rate $1/\tau_\downarrow \sim AQ^4 \sim \frac{m^*(k_BT)^2}{\hbar^3}\frac{d\sigma}{d\Omega}$.
The resulting equilibrium spin diffusivity is $D_s\sim v^2 \tau_\downarrow \sim \hbar^3/(m^{*2}k_BT\frac{d\sigma}{d\Omega})$.
In previous work, the same temperature dependence in terms of the equilibrium spin-drag relaxation rate was obtained for a mass-imbalanced system \cite{baarsma}.
Again, if there is instead a ``cloud''
of minority atoms, this will be their behavior when they are dilute enough to be nondegenerate.

It should be noted that the ``on-shell'' condition discussed in the beginning of the present paper is crucial in the analysis. As pointed out in Ref. \cite{josephson} in the context of heavy ion transport in He$^3$, if the definite effective mass assumption is not satisfied, the temperature dependence of transport coefficients can be significantly modified. We can understand this from the observation that the mass dependence of the scattering time is $\tau_\downarrow \sim m/m^*$. Therefore, when $m/m^* \ll 1$, the lifetime of the impurity can be small and thus its energy uncertainty becomes large and the assumption of elastic scattering breaks down. As a result, our analysis may not be applicable if the mass of the minority atom is much greater than that of the majority atom.

Next we discuss a general $d$-dimensional Fermi gas for $d\geq 2$. The argument is parallel to that given above for $d=3$.
The time evolution of the minority momentum distribution $f_d({\bf Q}_\downarrow, t)$ is the same as Eq.(\ref{Boltzmann_Eq})
except for the changes in dimension: $d^3{\bf q_\uparrow} \rightarrow d^d{\bf q_\uparrow}$ and $(2\pi)^3 \rightarrow (2\pi)^d$.
The solid angle $d\Omega$ now possesses $d$-dimensional hyperspherical form.
For $Q_\downarrow\ll k_{F\uparrow}$ the scattering cross section and relative speed remain essentially momentum independent and may be taken outside of the integral.
We use the same configuration of momenta (Fig.\ref{scheme}) and dimensionless parameters $p$, $\eta$, and $\gamma$ introduced above.
Once we fix the direction of ${\bf q}_\uparrow$, we are free to perform a rigid body rotation to the system of four momenta in $(d-2)$ angular directions.
Thus, $d\Omega ={\mathcal S}_{d-2} (\sin\eta)^{d-2} d\eta  \approx {\mathcal S}_{d-2}\eta^{d-2}d\eta  = {\mathcal S}_{d-2}(Q/k_{F\uparrow})^{d-1}\gamma^{d-2}d\gamma$ ($\eta$ is small).
${\mathcal S}_{d-2}$ is the surface area of a ($d-2$)-sphere, given by ${\mathcal S}_d = 2 \pi^{d/2}/\Gamma(d/2)$, where $\Gamma(x)$ is the gamma function.
The integration measure $d^d {\bf q}_\uparrow$ is scaled as $(m/m^*)k_{F\uparrow}^{d-2}Q^2 dp d\Omega_d$,
where $d\Omega_d$ is the solid angle element of ${\bf q}_\uparrow$ on the hypersphere.
We set the domain of both $\eta$ and $\theta$ to be $[0,\pi]$ for all $d\geq 2$.
Although $\eta$ and $\theta$ may be defined from $0$ to $2\pi$ for $d = 2$,
the reflection symmetry ensures us that we may integrate only from $0$ to $\pi$ and
the factor of $2$ can be absorbed in ${\mathcal S}_0 = 2$.
With all these scalings, we obtain
\begin{equation}\label{Scaled_Boltzmann_Eq_d}
\frac{df_d}{dt} = \frac{\hbar}{m^*}\left(\frac{d\sigma}{d\Omega}\right)_d\frac{{\mathcal S}_{d-2}}{(2\pi)^d}Q^{d+1}\int \gamma^{d-2}d\gamma d\Omega_d dp [...] ~,
\end{equation}
where we abbreviated the repeating phase factor $[n(\epsilon_{k_{\uparrow}})f_d(Q'_{\downarrow},t)(1 - n(\epsilon_{q_{\uparrow}}))
- n(\epsilon_{q_{\uparrow}})f_d(Q_{\downarrow},t)(1 - n(\epsilon_{k_{\uparrow}}))] = [...]$.
$(d\sigma/d\Omega)_d$ is the generalized differential scattering ``cross section" in $d$-dimensions.
Now we can explicitly see the scaling of the scattering rate is $\sim Q^{d+1}$ in a nonequilibrium regime and thus $\sim T^{(d+1)/2}$ near thermal equilibrium.
The typical distance traveled $l$ is $\sim t^{d/(d+1)}$ and
the volume in position-momentum configuration space that the minority atom occupies increases as $(l Q_\downarrow)^d \sim t^{d(d-1)/(d+1)}$.
The growth rate of the number of scattering steps remains $\sim \log t$.

Assuming isotropy, we can replace $d\Omega_d$ in Eq.(\ref{Scaled_Boltzmann_Eq_d}) with ${\mathcal S}_{d-2}(\sin\theta)^{d-2}d\theta$.
As we did for $d=3$, we multiply $\int dQ_\downarrow' \delta(Q'_\downarrow - \sqrt{Q_\downarrow^2 + \gamma^2 Q^2 - 2Q_\downarrow Q \gamma \cos\theta})$. Then, for an odd $d$ we do the $\theta$ integral first, while for an even $d$ we do the $\gamma$ integral first.
This gives
\begin{eqnarray}
&&\frac{df_d(Q_\downarrow, t)}{dt} = \frac{\hbar}{m^{*}}\left(\frac{d\sigma}{d\Omega}\right)_d\frac{({\mathcal S}_{d-2})^2}{(2\pi)^d}\int dp [...]\times\nonumber\\
&&\int\frac{Q^3Q'_\downarrow dQ'_\downarrow d\gamma}{2^{d-3}Q^{d-2}_\downarrow}\left[(2Q_\downarrow Q \gamma)^2 - (Q_\downarrow^2 + \gamma^2 Q^2 - Q'^2_\downarrow)^2\right]^{\frac{d-3}{2}}\nonumber\label{Isotropic_Boltzmann_Eq_odd_d} \\
\\
&&\text{or}\nonumber\\
&&\int\frac{Q^2 Q'_\downarrow dQ'_\downarrow d\theta}{\sqrt{Q'^2_\downarrow - Q^2_\downarrow \sin^2\theta}} \gamma_0^{d-2}(\sin\theta)^{d-2},\label{Isotropic_Boltzmann_Eq_even_d}
\end{eqnarray}
where Eq.(\ref{Isotropic_Boltzmann_Eq_odd_d}) is for an odd $d$ and Eq.(\ref{Isotropic_Boltzmann_Eq_even_d}) is for an even $d$.
$\gamma_0$ is the value of $\gamma(\theta)$ that satisfies the delta function
(it has two roots and which root to use depends on the magnitudes of $Q_\downarrow$, $Q'_\downarrow$, and $\theta$).
For an odd $d$, the integration range of $\gamma$ is from $|Q_\downarrow - Q'_\downarrow|/Q$ to $(Q_\downarrow + Q'_\downarrow)/Q$
and the integrand is a polynomial. Therefore, the $\gamma$ integral is elementary,
although a formula for a general odd $d$ is fairly lengthy so we do not present here explicitly.
On the other hand, the $\theta$ integral for an even $d$ is not elementary so
it may have to be evaluated numerically.

We introduce $A_d = \frac{\hbar}{m^{*}}\left(\frac{d\sigma}{d\Omega}\right)_d\frac{({\mathcal S}_{d-2})^2}{(2\pi)^d}$,
which has dimensions of (length$^{(d+1)}$/time).
When $d = 3$, $A_d = A = \frac{\hbar}{2\pi m^*}\frac{d\sigma}{d\Omega}$, as above.
At zero temperature, we can do the integral $\int dp [...]$ to obtain
\begin{eqnarray}\label{Boltzmann_Eq_d}
&&\frac{f_d(Q_\downarrow, t)}{dt} = \\
&&A_d\left[\int_{Q_\downarrow}^\infty dQ'_\downarrow h_d(Q'_\downarrow, Q_\downarrow)f_d(Q'_\downarrow,t) - C_d Q^{d+1}_\downarrow f_d(Q_\downarrow, t)\right]~.\nonumber
\end{eqnarray}
The function $h_d(Q'_\downarrow, Q_\downarrow)$ and the constant $C_d$ should be determined by Eq.(\ref{Isotropic_Boltzmann_Eq_odd_d}) or Eq.(\ref{Isotropic_Boltzmann_Eq_even_d}) and the $p$ integral.
For an odd $d$, $C_d$ can be found exactly, while for an even $d$, $C_d$ could be found numerically.
As we saw earlier, for $d=3$, $C_3 = 2/15$ and
$h_3(Q'_\downarrow, Q_\downarrow) = Q'_\downarrow(Q'^2_\downarrow - Q^2_\downarrow)$.

Before we scale the long time dynamics of Eq.(\ref{Boltzmann_Eq_d}), let us estimate the spin diffusivity in $d$ dimensions.
Near thermal equilibrium, $v^2 \sim (\hbar Q/m^*)^2 \sim (k_B T)/ m^*$ and the scattering time $\tau$ is $\sim 1/(A_d Q^{d+1})$.
Thus, the spin diffusivity in $d$ dimensions is $D_s \sim v^2\tau \sim \frac{\hbar^d}{m^{*(d+1)/2} (k_B T)^{(d-1)/2} (d\sigma/d\Omega)_d}$
when the minority gas is nondegenerate and the majority gas is an ideal Fermi sea or a Fermi liquid.

It is straightforward to obtain a scaling form once we introduce the following scaling of the momentum and the distribution:
\begin{eqnarray}
s_d &=& (A_d t)^{1/(d+1)}Q_\downarrow\\
f_d(Q_\downarrow, t) &=& (A_d t)^{d/(d+1)} g_d(s_d) ~.
\end{eqnarray}
The scaled integro-differential equation is,
\begin{eqnarray}
&&\frac{d}{d+1}g_d(s_d) + \frac{s_d}{d+1}\frac{d g_d(s_d)}{d s_d} \nonumber\\
&=& \int^{\infty}_{s_d} d s'_d g_d(s'_d) h(s'_d, s_d)  - C_d s_d^{d+1} g_d(s_d)~.
\end{eqnarray}
We see that the asymptotic form of $g_d(s_d)$ at large $s_d$ is $\sim \exp[-C_d s^{d+1}_d]$.

For $d=2$, which is
the other experimentally accessible case:
$A_2 = \frac{\hbar}{\pi^2 m^*}\left(\frac{d\sigma}{d\Omega}\right)_2$,
$C_2 \cong 0.45$ and
$h_2(Q'_\downarrow, Q_\downarrow) = \int^\pi_0 d\theta \frac{Q'_\downarrow (Q'^2_\downarrow - Q^2)}{2\sqrt{Q'^2_\downarrow - Q^2_\downarrow\sin^2\theta}}$.
The spin diffusivity near thermal equilibrium in $d=2$ is
$D_s \sim \frac{\hbar^2}{m^{*3/2}(k_B T)^{1/2}(d\sigma/d\Omega)_2}$.

So far, our calculation relied heavily upon the phase space integral.
In $d = 1$, however, the reduced phase space greatly restricts the scattering process.
The scattering of an impurity from a $T=0$ Fermi sea for $Q_\downarrow \ll k_{F\uparrow}$ via creating a single particle-hole pair is forbidden in $d=1$ by
energy and momentum conservation unless $m^* \ll m$, which is not true in our case.
Therefore, a low-momentum moving impurity in $d=1$ is stable at zero temperature against scattering at this order.
This is why the results of the present paper only apply for $d\geq 2$.
The dynamics of a heavy impurity in a one-dimensional Luttinger liquid
due to other higher-order scattering processes is discussed in Ref. \cite{castro}.

Finally, we consider an experimental procedure to possibly observe the expected superdiffusive behavior. 
Initially, the majority and impurity atoms are trapped by species-selective potentials and cooled to very low $T$. 
The impurity atoms are tightly trapped at the center of a majority cloud, and thus they are initially degenerate, but with a lower Fermi momentum than the
majority atoms. (Here we discuss the case where the impurity atoms are fermions; if the impurity atoms are bosons,
the specifics of how to set up an appropriate initial condition will be different).
Then, release the impurity atoms but not the majority atoms and observe the subsequent expansion of the minority cloud using {\it in situ} imaging.
At first, the impurity cloud should expand ballistically since scattering is highly restricted by both majority and minority Pauli blocking.
Quickly the impurity cloud expands and becomes nondegenerate and superheated relative to the majority gas.  
At this point the impurities start doing the superdiffusive motion discussed in this paper, which should be reflected in a superdiffusive expansion of the minority cloud.  This superdiffusive motion persists until the impurity atoms cool to
near the temperature of the majority Fermi gas. 
Then the impurity motion crosses over to standard diffusion.  Recently, the expansion dynamics of initially localized impurities in a 1-dimensional Bose gas has
been realized and imaged \cite{inguscio}.  

In conclusion, we have studied the nonequilibrium and near-equilibrium motion of nondegenerate impurity atoms in a low-temperature Fermi sea.
In the nonequilibrium regime we analyzed the unusual superdiffusive random geometric walk performed by the impurity atom for systems
with dimensionality $d\geq 2$.  At nonzero temperature this crosses over to standard diffusion at long times.
We find that the equilibrium spin diffusivity $D_s\sim T^{-(d-1)/2}$ in this regime where the majority atoms
are degenerate, while the minority atoms are not.  This temperature dependence of the spin diffusivity is
something that may be measured soon in experiments on 2- and 3-dimensional Fermi gases.  The superdiffusive
nonequilibrium behavior will be more of a challenge to explore experimentally, since it requires a
hierarchy of three energy scales: $k_BT\ll\frac{\hbar^2Q^2_\downarrow}{2m^*}\ll k_BT_{F\uparrow}$, and thus
a very cold majority Fermi gas.

We thank Marco Schiro for many discussions. We thank Martin Zwierlein, Jeewoo Park, and Ariel Sommer for discussions about experiments. H. K. also thanks Sungjin Oh for help and Colin Parker for suggestions. This work is supported by ARO Award W911NF-07-10464 with funds from the DARPA OLE Program. H. K. is partially supported by Samsung Scholarship.


\begin{thebibliography}{99}
\bibitem{mitra}
T. K. Mitra, A. Chatterjee and S. Mukhopadhyay, Phys. Rep. {\bf 153}, 91 (1987).

\bibitem{devreese}
J. T. Devreese and A. S. Alexandrov, Rep. Prog. Phys. {\bf 72}, 066501 (2009).

\bibitem{abe}
R. Abe and K. Aizu, Phys. Rev. {\bf 123}, 10 (1961).

\bibitem{clark}
R. C. Clark, Proc. Phys. Soc. (London) {\bf 82}, 785 (1963).

\bibitem{schappert}
G. T. Schappert, Phys. Rev. {\bf  168}, 162 (1968).

\bibitem{zwierlein}
M. W. Zwierlein, A. Schirotzek, C. H. Schunck and W. Ketterle, Science {\bf 311}, 492 (2006).

\bibitem{partridge1}
G. B. Partridge, W. Li, R. I. Kamar, Y. Liao and R. G. Hulet, Science {\bf 311}, 503 (2006).

\bibitem{shin1}
Y. Shin, M. W. Zwierlein, C. H. Schunck, A. Schirotzek and W. Ketterle, Phys. Rev. Lett. {\bf 97}, 030401 (2006).

\bibitem{partridge2}
G. B. Partridge, W. Li, Y. A. Liao, R. G. Hulet, M. Haque and H. T. C. Stoof, Phys. Rev. Lett. {\bf 97}, 190407 (2006).

\bibitem{shin2}
Y. Shin, C. H. Schunck, A. Schirotzek and W. Ketterle, Nature {\bf 451}, 689 (2008).

\bibitem{shin3}
Y. I. Shin, Phys. Rev. A {\bf 77}, 041603 (2008).

\bibitem{nascimbene}
S. Nascimb${\rm \grave{e}}$ne, N. Navon, K. J. Jiang, F. Chevy and C. Salomon, Nature {\bf 463}, 1057 (2010).

\bibitem{chevy}
F. Chevy, Phys. Rev. A {\bf 74}, 063628 (2006).

\bibitem{lobo}
C. Lobo, A. Recati, S. Giorgini and S. Stringari, Phys. Rev. Lett. {\bf 97}, 200403 (2006).

\bibitem{combescot}
R. Combescot, A. Recati, C. Lobo and F. Chevy, Phys. Rev. Lett. {\bf 98}, 180402 (2007).

\bibitem{prokofev}
N. Prokof'ev and B. Svistunov, Phys. Rev. B {\bf 77}, 020408 (2008).

\bibitem{mora}
C. Mora and F. Chevy, Phys. Rev. Lett. {\bf 104}, 230402 (2010).

\bibitem{bruun}
G. M. Bruun, A. Recati, C. J. Pethick, H. Smith and S. Stringari, Phys. Rev. Lett. {\bf 100}, 240406 (2008).

\bibitem{sommer2}
A. Sommer, M. Ku and M. W. Zwierlein, New J. Phys. {\bf 13}, 055009 (2011).

\bibitem{trenkwalder}
A. Trenkwalder, C. Kohstall, M. Zaccanti, D. Naik, A. I. Sidorov, F. Schreck and R. Grimm, Phys. Rev. Lett. {\bf 106}, 115304 (2011).

\bibitem{schirotzek}
A. Schirotzek, C-H. Wu, A. Sommer and M. W. Zwierlein, Phys. Rev. Lett. {\bf 102}, 230402 (2009).

\bibitem{baarsma}
J. E. Baarsma, J. Armaitis, R. A. Duine and H. T. C. Stoof, arXiv:1110.3143v1.

\bibitem{josephson}
B. D. Josephson and J. Lekner, Phys. Rev. Lett. {\bf 23}, 111 (1969).

\bibitem{castro}
A. H. Castro Neto and M. P. A. Fisher, Phys. Rev. B {\bf 53}, 9713 (1996).

\bibitem{inguscio}
J. Catani, G. Lamporesi, D. Naik, M. Gring, M. Inguscio, F. Minardi, A. Kantian and T. Giamarchi, Phys. Rev. A {\bf 85}, 023623 (2012).

\end{thebibliography}
\end{document}